\documentclass[aps,pre,preprint,groupedaddress,showpacs]{revtex4}
\usepackage{graphicx}
\usepackage{amsmath}
\begin{document}
\title{On the mechanism of branching in negative ionization fronts}
\author{Manuel Array\'as$^{1}$, Marco A. Fontelos$^{2}$ and Jos\'e L. Trueba$^{1}$}
\affiliation{$^{1}$Departamento de Matem\'aticas y F\'{\i}sica
Aplicadas y Ciencias de la Naturaleza, Universidad Rey Juan
Carlos, Tulip\'{a}n s/n, 28933 M\'{o}stoles, Madrid, Spain}
\affiliation{$^{2}$Departamento de Matem\'aticas, Universidad
Aut\'onoma de Madrid, 28049 Cantoblanco, Madrid, Spain}

\begin{abstract}
We explain a mechanism for branching of a planar negative front. Branching occurs as the result of a balance between the destabilizing effect of impact ionization and the stabilizing effect of electron diffusion on ionization fronts. The dispersion relation for transversal perturbation is obtained analytically and reads: $s = |k|/[2 ( 1 + |k|)] - D |k|^2$, where $D$, which is assumed to be small, is the ratio between the electron diffusion coefficient and the intensity of the externally imposed electric field. We estimate the spacing $\lambda$ between streamers in a planar discharge and deduce a scaling law $\lambda \sim D^{1/3}$.
\end{abstract}
 
\date{\today}
\maketitle

One of the greatest unsolved problems in the physics of electric discharges is the clarification of the mechanism of branching. When a strong electric field is applied to a non-conducting medium such as a gas, narrow channels of ionized matter called streamers may be formed. The phenomena has been observed in a wide range of scales, from small-gap discharges of few centimetres up to scales of kilometres such as in sprites discharges from a thunder cloud. The pattern of this branching resembles the ones observed in the propagation of cracks, dendritic growth and viscous fingering. Those phenomena are known to be governed by deterministic equations rather than by stochastic events. In this paper, for the first time, it is derived a quantitative prediction of the branching characteristic length based on a deterministic model. The results give an explicit dependence of branching with the electric field and the pressure of the gas as it has been observed qualitatively in experiments.         

We use a fluid approximation to describe electric breakdown in
non-attaching gases such as nitrogen. In these gases, there are
indications showing that the most important source of electrons
and positive ions from neutral molecules is impact ionization
\cite{liu}. This is the only process that will be taken into
account in this work. This leads to a minimal streamer model from
which the electron $N_{e}$ and positive ion $N_{p}$ densities can
be obtained. The minimal model reads
\begin{eqnarray}
\label{elect} \frac{\partial N_{e}}{\partial \tau} &=& \nabla
\cdot \left( N_{e} {\boldsymbol{\cal E}} + D_{e} \, \nabla N_{e}
\right) + N_{e}
|{\boldsymbol{\cal E}}| e^{-1/|{\boldsymbol{\cal E}}|},\\
\label{ion} \frac{\partial N_{p}}{\partial \tau} &=& N_{e}
|{\boldsymbol{\cal E}}| e^{-1/|{\boldsymbol{\cal E}}|}, \\
\label{gauss} N_{p} - N_{e} &=& \nabla \cdot {\boldsymbol{\cal
E}},
\end{eqnarray}
in which ${\boldsymbol{\cal E}}$ is the local electric field in
the gas and $D_{e}$ is the electronic diffusion coefficient. The
first equation means that the electron density varies in a point
as a result of (i) the electric current contribution $\nabla (
N_{e} {\boldsymbol{\cal E}})$, (ii) the electronic diffusion term
$D_{e} \, \nabla^2 N_{e}$, and (iii) the impact ionization source term
$N_{e} |{\boldsymbol{\cal E}}| e^{-1/|{\boldsymbol{\cal E}}|}$.
The second equation means that the positive ion density in a point
of the gas only varies in time as a result of impact ionization,
since the mobility of ions is much smaller than that of the
electrons. The third equation is Poisson's law for the electric
field, which is supposed to be irrotational since the magnetic
effects are neglected.

All the quantities in Eqs.~(\ref{elect})--(\ref{gauss})
are dimensionless. The scales of these quantities for nitrogen
\cite{montjin} depend on the gas pressure $p$. The characteristic
field is ${\cal E}_{0} = 200 \, \mbox{kV}/\mbox{cm} \,
(p/1\,\mbox{bar})$. The characteristic length is $R_{0} = 2.3 \,
\mu\mbox{m} \, (p/1\,\mbox{bar})^{-1}$. The characteristic
velocity is $U_{0} = 76 \times 10^{6} \, \mbox{cm}/\mbox{s}$. The
characteristic time is $\tau_{0} = 3 \, \mbox{ps} \,
(p/1\,\mbox{bar})^{-1}$. The characteristic particle density is
$N_{0} = 4.7 \times 10^{14} \, \mbox{cm}^{-3} \,
(p/1\,\mbox{bar})^2$. The characteristic diffusion coefficient is
$D_{0} = 1.8 \times 10^{4} \, \mbox{cm}^2/\mbox{s} \,
(p/1\,\mbox{bar})^{-1}$.

In the minimal model, a spontaneous branching of negative
streamers has been observed numerically \cite{ME}, as it occurs in
experimental situations \cite{Pasko}. In order to understand this
branching, the dispersion relation for transversal Fourier-modes
of planar negative shock fronts (without diffusion) has been
derived \cite{ME1}. For perturbations of small wave number $k$,
the planar shock front becomes unstable with a linear growth rate
proportional to $k$, but all the modes with large enough wave
number $k$ seem to grow at the same rate.

In this paper, we address the problem of the selection of a
particular wave number in the perturbation of negative planar
fronts. We will obtain a new dispersion relation depending
explicitly on the electric field and the electronic diffusion
coefficient (which depends also on the gas pressure). Our analysis
will show that the electron density $N_{e}$ may develop steep
fronts of thickness $O(\sqrt{D_{e}})$, satisfying an equation
analogous to Fisher equation \cite{Kolmogorov}. A surprising fact
established during the last 30 years is that the combination of
sharp interfaces with small diffusive effects may result in
asymptotic limits (for $D_{e}\ll 1$) in which the motion of the
interface is described by equations involving solely geometrical
properties such as its mean curvature \cite{Allen}. These analysis
concerned a model today known as Allen-Cahn equation. Subsequent
work \cite{Keller} showed that the points of the interface
separating two different phases move along the normal direction with a
velocity proportional to its mean curvature. This kind of dynamics
is termed "mean curvature flow". Many mathematicians have
contributed to provide a rigorous proof of the convergence of
Allen-Cahn model to motion by mean curvature
\cite{Cahn,Alikakos,Hilhorst}. Remarkably, some of these limiting
models may be such that the solutions develop branch-like
patterns. In this work, we exploit some of these ideas to study
the motion of ionization fronts. We will show
that a planar front separating a (partly) ionized region from a
region without charge is such that small geometrical perturbations
in the charge distribution lead to a motion of the front affected
by two opposed effects: electrostatic repulsion of electrons and
electron diffusion. The first effect tends to destabilize the
front while the second acts effectively as a mean curvature
contribution to the velocity of the front, thus stabilizing it.
The net result is the appearance of fingers with a characteristic
thickness determined by the balance of these two opposing actions.

In order to study the evolution and branching of ionization fronts, we
consider the following experimental situation. The space between
two large planar plates, situated at $x=0$ and $x=d$ respectively
($x$ is the vertical axis and we suppose that $d \gg 1$), is
filled with a non-attaching gas like nitrogen. A stationary
electric potential difference is applied to these plates, so that
an electric field is produced in the gas. The initial electric
field is directed from the anode to the cathode, along the
negative $x$ axis, and is uniform in the space between the plates.
To initiate the avalanche, an initial seed of ionization is set
near the cathode. We first study the evolution of planar negative
ionization fronts towards the anode.

We will concentrate in the study of the dynamics under the effect
of strong external electric fields. We denote the modulus of
the dimensionless electric field at large distance from the
cathode as ${\cal E}_{\infty }$ and we will assume that ${\cal
E}_{\infty} \gg 1$. Under these circumstances, it is natural to
rescale the dimensionless quantities in the minimal model as
${\boldsymbol{\cal E}} = {\cal E}_{\infty } \, {\bf E}$, $N_{e} =
{\cal E}_{\infty } \, n_{e}$, $N_{p} = {\cal E}_{\infty } \,
n_{p}$, and $\tau = t /{\cal E}_{\infty }$. For ${\cal E}_{\infty}
\gg 1$, this system can be approximated by
\begin{eqnarray}
\frac{\partial n_{e}}{\partial t}-\nabla \cdot \left( n_{e} {\bf
E} + D\,\nabla n_{e} \right) &=& n_{e} |{\bf E}| ,  \label{ec1} \\
\frac{\partial n_{p}}{\partial t} &=& n_{e} |{\bf E}| ,  \label{ec2} \\
\nabla \cdot {\bf E} &=& n_{p} - n_{e}, \label{ec3}
\end{eqnarray}
where $D = D_{e}/{\cal E}_{\infty}$ is, in general, a small
parameter. Our approximation will be valid in all regions where
$|{\bf E}| = O(1)$. These are the regions of interest since the
electric field is not expected to vary much in the neighbourhood of
the ionization front and we will show that it is in this region
where the mechanisms leading to branching take place.

In the evolution of the ionization wave along the $x$ axis, the
rescaled electric field can be written as ${\bf E} = E {\bf
u}_{x}$ where $E < 0$, so that $|{\bf E}| = |E| = -E$. It is very
simple to compute travelling wave solutions when $D=0$. It can be
shown \cite{prelargo} that these solutions exist for any $c \geq
1$. We will be interested in the limit $c\rightarrow 1$ since it
is well known that a compactly supported initial data
(representing a seed of ionization located in some region)
develops fronts travelling with this velocity. In the case $c=1$
the solution can be obtained straightforwardly, giving (with $\xi
=x-ct$ and $c=1$)
\begin{equation}
E(\xi)=\left\{
\begin{array}{c}
- e^{\xi}, \, \, \mbox{for} \,\, \xi <0  \\
- 1, \, \, \mbox{for} \,\, \xi \geq 0
\end{array}
\right. , \, \, n_{e} (\xi )=\left\{
\begin{array}{c}
1, \, \, \mbox{for} \,\, \xi <0 \\
0, \, \, \mbox{for} \,\, \xi \geq 0
\end{array}
\right. , \, \, n_{p} (\xi )=\left\{
\begin{array}{c}
1-e^{\xi}, \, \, \mbox{for} \, \, \xi <0 \\
0, \,\, \mbox{for} \,\, \xi \geq 0
\end{array}
\right. . \label{anst7}
\end{equation}
In the case $0 < D \ll 1$, it is known \cite{Ute} that all initial data
decaying at infinity faster than $Ae^{-\lambda ^{\ast }x}$, with
$\lambda ^{\ast }=2/\sqrt{D}$, will develop travelling waves with
velocity $c=1+2\sqrt{D}$. If $D \ll 1$, the profiles for $n_{p}$
and $E$ will vary very little from the profiles with $D=0$. On the
other hand, $n_{e}$ will develop a boundary layer at the front
smoothing the jump from $n_{e} =1$ to $n_{e} = 0$. Approximating
at the boundary layer $n_{p} = 0$, $E = -1$, we obtain the
equation
\begin{equation}
-2 \frac{\partial n_{e}}{\partial \chi} - \frac{\partial^2
n_{e}}{\partial \chi^2} = n_{e}(1-n_{e}) \label{fisher}
\end{equation}
where $\chi = (x-(1+2\sqrt{D}) t)/ \sqrt{D}$, together with the
matching conditions $n_{e}(-\infty )=1$ and $n_{e}(+\infty )=0$.
Eq.~(\ref{fisher}) is the well known equation for the
travelling waves of Fisher's equation. It appears in the context
of mathematical biology \cite{Murray} and is known to have
solutions subject to our matching conditions. This means that we
have a boundary layer of width $\sqrt{D}$ at $\xi =0$ in which Eq.~(\ref{fisher}) gives the solution for the electron
density $n_{e}$. Before this layer, we have $n_{e} \approx 1$, and
after the layer, $n_{e} \approx 0$. When $D = 0$, this is the
shock front of Eq.~(\ref{anst7}). In what respect to $n_{p}$
at the boundary layer, at first order in $D$ one obtains
\begin{equation}
n_{p}(\chi )= - \sqrt{D} \int_{\chi }^{\infty } n_{e}(\chi )d\chi,
\label{dpeq10}
\end{equation}
so that $\partial n_{p}/\partial x$ is $O(1)$ at the boundary
layer.

\begin{figure}
\centering
\includegraphics[width=0.45\textwidth,height=0.35\textwidth]{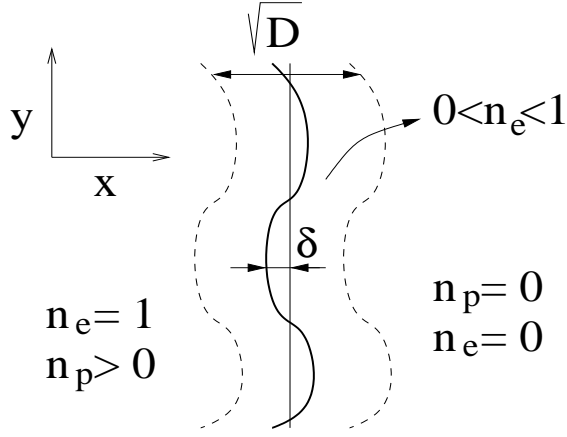}
\caption{A planar front is perturbed by a displacement of characteristic size $\delta$. The boundary layer has a characteristic size $\sqrt{D}$ as explained in the text. In this boundary layer, the electron density $n_e$ goes from 1 to 0 and the positive ion density $n_p$ decays to 0. The front moves to the right along the positive $x$ axis.} \label{fig1}
\end{figure}

Now we make a perturbation in the transversal direction $y$. We introduce a new system of coordinates in the form $\overline{t} = t$,
$\overline{y} = y$, $\overline{x} = x - \delta \, \varphi (x,y,t)$
so that, at $t=0$, $n_{e}^{(0)}(\overline{x}), n_{p}^{(0)}(\overline{x}),$ and
$E^{(0)}(\overline{x})$ correspond to the profiles of the
travelling wave computed in the previous paragraph, and $\delta$
is a sufficiently small parameter (see Fig.~\ref{fig1}). By
doing this, we follow a strategy analogous to the one used in
Rubinstein et al.~\cite{Keller} to deduce the asymptotic
approximation of Allen-Cahn equation by mean curvature flow.

We introduce the perturbed electric field and electron and ion
densities as
\begin{eqnarray}
{\bf E} &=& E^{(0)}\,{\bf u}_{x} + \delta \left( E^{(1)}_{x}
\,{\bf u}_{x} + E^{(1)}_{y} {\bf u}_{y} \right) , \label{pertelec} \\
n_{e} &=& n_{e}^{(0)} + \delta^{2} \, n_{e}^{(1)} , \label{pertsigma} \\
n_{p} &=& n_{p}^{(0)} + \delta \, n_{p}^{(1)} , \label{pertrho}
\end{eqnarray}
and we will select the function $\varphi (x,y,t)$ and the
$O(\delta)$ corrections $E^{(1)}_{x},E^{(1)}_{y},n_{p}^{(1)}$ in
such way that Eqs.~(\ref{pertelec})--(\ref{pertrho}) represent correct series expansions.
Specifically, the correction to the travelling wave profiles for
$n_{e}$ will be $O(\delta^2)$.

We insert these expressions into Eqs.~(\ref{ec1})--(\ref{ec3}). Then we impose that $O(\delta ^{0})$
terms and $O(\delta ^{1})$ terms vanish. The solution of the
equations at order $O(\delta^{0})$ is the travelling wave found
previously, so that $n_{e}^{(0)} (\overline{x},\overline{t}) =
n_{e}^{(0)} (\overline{x} - \overline{t})$ is given by Fisher's
equation (\ref{fisher}), $n_{p}^{(0)} (\overline{x} -
\overline{t})$ is given by Eq.~(\ref{dpeq10}), and
$E^{(0)} (\overline{x} - \overline{t})$ is the solution of
Poisson's equation corresponding to these particle densities.

At order $O(\delta^{1})$, after some manipulations, and taking
into account that $\partial /\partial \overline{x}=O (D^{-1/2})$
and $\partial n_{p}^{(0)}/\partial \overline{x} =O(1)$ at the boundary layer, and
that $D$ is a small parameter, the equation for the perturbed ion
density $n_{p}^{(1)}$ decouples from the equations for the
perturbed electron density $n_{e}^{(1)}$ and the perturbed
electric field $(E_{x}^{(1)}, E_{y}^{(1)})$. The system at $O(\delta ^{1})$ order is then given by the evolution equations
\begin{eqnarray}
\frac{\partial \varphi}{\partial \overline{t}} + E^{(1)}_{x} - D
\nabla_{(\overline{x},\overline{y})}^2 \varphi  -  2 D
\frac{\partial \varphi}{\partial \overline{x}} \frac{\partial^2
n_{e}^{(0)} /
\partial \overline{x}^{2}}{\partial n_{e}^{(0)} / \partial
\overline{x}} &=& 0 , \label{e9} \\
\frac{\partial E^{(1)}_{x}}{\partial \overline{x}} +
\frac{\partial E^{(1)}_{y}}{\partial \overline{y}} -
\frac{\partial \varphi}{\partial \overline{x}} \left( n_{p}^{(0)}
- n_{e}^{(0)} \right) &=& 0. \label{e9b}
\end{eqnarray}
Observe that the system (\ref{e9})--(\ref{e9b}) simplifies if one assumes that $\varphi$ is independent of $\overline{x}$. This is a valid
assumption at least for a short period of time (the one in which
stability is analyzed) if one assumes $\varphi$ independent of
$\overline{x}$ at $t=0$.

It is more convenient to formulate Eq.~(\ref{e9b}) in terms
of the electric potential. We note that the total electric field
has to be irrotational since the magnetic field is negligible. So
we will assume that ${\bf E} = - \nabla V$, where $V$ is an
electric potential that can be written as
$V(\overline{x},\overline{y}) = V^{(0)}(\overline{x}) + \delta \,
V^{(1)}(\overline{x}, \overline{y})$. At order $O( \delta^{0})$,
Poisson's equation implies that $V^{(0)}(\overline{x})$ is an
electric potential associated to the electric field
$E^{(0)}(\overline{x})$. At order $O( \delta^{1})$, Poisson's
equation implies that $V^{(1)}$ satisfies
\begin{equation}
- \nabla_{(\overline{x},\overline{y})}^2
V^{(1)}(\overline{x},\overline{y}) = -
\nabla_{(\overline{x},\overline{y})}^2 \varphi \frac{\partial
V^{(0)}(\overline{x})}{\partial \overline{x}} - 2 \frac{\partial
\varphi}{\partial \overline{x}} \frac{\partial^2
V^{(0)}(\overline{x})}{\partial \overline{x}^2} + n_{p}^{(1)} ,
\label{potential1}
\end{equation}
with the condition of decaying at $|\overline{x}| \rightarrow
\infty $. Again, by noting $\partial /\partial \overline{x}=O
(D^{-1/2})$, we can neglect $n_{p}^{(1)}$ in Eq.~(\ref{potential1}),
and solve the resulting equation by taking Fourier transform in
$\overline{y}$ to find the following value for the Fourier
transform of $E^{(1)}_{x}$,
\begin{equation}
\hat{E}^{(1)}_{x} (\overline{x},k) = - \frac{|k| \hat{\varphi}
(k)}{2} \times \left\{
\begin{array}{c}
\frac{1}{1+ |k|} e^{-|k| \overline{x}} , \, \, \mbox{for} \,\, \overline{x} \geq 0 \\
\frac{-2|k|}{1 - |k|^{2}} e^{\overline{x}} + \frac{1}{1 - |k|}
e^{|k| \overline{x}}, \, \, \mbox{for} \,\, \overline{x} \leq 0
\end{array}
\right. . \label{potential10}
\end{equation}
The front is at a neighbourhood of $O(D^{1/2})$ width around
$\overline{x} =0$. While $| k | \ll D^{-1/2}$, the exponentials in
Eq.~(\ref{potential10}) can be neglected in this region, and we can
write
\begin{equation}
\hat{E}^{(1)}_{x} (0,k) = - \frac{|k| \hat{\varphi} (k)}{2 (1 +
|k|)} , \label{potential11}
\end{equation}
i.e. a field independent of $\overline{x}$.

Assuming that $\varphi$ does not depend on $\overline{x}$, we can
write Eq.~(\ref{e9}) in the form
\begin{equation}
\frac{\partial \varphi}{\partial \overline{t}} + E^{(1)}_{x} - D
\frac{\partial^2 \varphi}{\partial \overline{y}^2} = 0 .
\label{buena1}
\end{equation}
The third term at the right hand side of Eq.~(\ref{buena1}) is the contribution of the electron diffusion to the evolution of the front and is also its linearized mean curvature. Taking Fourier transform of Eq.~(\ref{buena1}) in $\overline{y}$, and using Eq.~(\ref{potential11}), we find
\begin{equation}
\frac{\partial \hat{\varphi} (k)}{\partial \overline{t}} -
\frac{|k| \hat{\varphi} (k)}{2 (1+|k|)} + D|k|^2 \hat{\varphi} (k)
= 0. \label{buena2}
\end{equation}
Now, let us write the following ansatz for $\hat{\varphi}$,
\begin{equation}
\hat{\varphi} (k,\overline{t})=e^{s \overline{t}} \hat{\phi} (k).
\label{buena3}
\end{equation}
Introducing this expression into Eq.~(\ref{buena2}), we
obtain the relation
\begin{equation}
s = \frac{|k|}{2 ( 1 + |k|)} - D |k|^2 , \label{muybuena1}
\end{equation}
that gives the dispersion curve of the transversal perturbations
of the planar negative ionization front explicitly in terms of the
parameter $D = D_{e}/{\cal E}_{\infty}$ (see Fig.~\ref{fig2}). From this result we can obtain some important consequences on the branching of streamers:

\begin{figure}
\centering
\includegraphics[width=0.45\textwidth,height=0.35\textwidth]{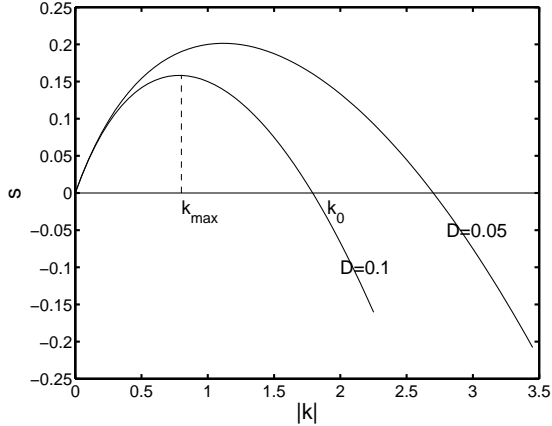}
\caption{The dispersion relation given by Eq.~(\ref{muybuena1}) for $D=0.1$ and $D=0.05$. For the first curve, the position of the maximum ($k_{max}$) and the value where the curve changes sign ($k_0$) are shown. The physical roles of these values are discussed at the text.} \label{fig2}
\end{figure}

(i) There exists a maximum of $s (|k|)$ that selects the
wavelength of the perturbation. When $D$ is a small parameter,
this maximum is approximately located at
\begin{equation}
k_{max} \approx \left( \frac{{\cal E}_{\infty}}{4 D_{e}}
\right)^{1/3} . \label{muybuena2}
\end{equation}
Notice that $k_{max}$ is $O(D^{-1/3})$, so that $|k| \overline{x}$
can be safely approximated by zero in the boundary layer. This
justifies the assumption, done previously in this work, that
$E_{x}^{(1)}$, and hence $\varphi$, are independent of
$\overline{x}$ at this order. The value of $k_{max}$ corresponds
to a typical spacing between fingers given by
\begin{equation}
\lambda_{max} = \frac{2 \pi}{k_{max}} \approx 10 \left(
\frac{D_{e}}{{\cal E}_{\infty}} \right)^{1/3} . \label{muybuena4}
\end{equation}
This is an equation in dimensionless units. If we introduce
typical scales for nitrogen (see the paragraph below Eq.~(\ref{gauss})), then we can obtain the typical distance between
two consecutive branches in a negative streamer discharge. This
distance will decrease when the gas pressure or the initial
electric field in the gas increases.

(ii) There exists an stability threshold, i.e.~a value of $k$ delimiting the stability region, which corresponds to the
nontrivial zero of the function $s(k)$,
\begin{equation}
k_{0} \approx \sqrt{\frac{{\cal E}_{\infty}}{2D_{e}}}.
\label{muybuena6}
\end{equation}
The wavelength $\lambda_{0}$ associated to this wave number is
\begin{equation}
\lambda_{0} = \frac{2 \pi}{k_{0}} \approx 8.9 \sqrt{
\frac{D_{e}}{{\cal E}_{\infty}}}. \label{muybuena7}
\end{equation}

Although the predictions are made for negative planar fronts, they agree with the observed fact that, for positive discharges, the number of streamers increases with the electric field and the pressure. Those effects are accounted by the expression (\ref{muybuena4}). 

We can now provide a qualitative mechanism of streamer branching. When the radius of a streamer becomes larger than the critical length given by Eq.~(\ref{muybuena7}), the streamer becomes unstable and branching develops. The electric field which should be taken into Eq.~(\ref{muybuena4}) in case of inhomogeneous electric discharges is the local field at the front of the streamer. In order to test the predictions, experimental evidence should be provided in the range where the approximations of large electric fields and small diffusion coefficient are valid. 

To conclude, we have derived analytically the characteristic length of branching for planar negative ionization fronts. At the same time, this prediction can be considered as a test for the validity of the minimal deterministic model on which this calculation is based.

We thank Robert Deegan for useful discussions on this work.
This paper has been partially supported by the Spanish Ministry of
Science and Technology grant BFM2002-02042, and by the Universidad
Rey Juan Carlos grant PPR-2004-38.

\end{document}